\documentstyle[12pt]{article}
\def\Re{{\cal R \mskip-4mu \lower.1ex \hbox{\it e}\,}}
\def\Im{{\cal I \mskip-5mu \lower.1ex \hbox{\it m}\,}}
\def\ie{{\it i.e.}}
\def\eg{{\it e.g.}}

\def\etal{{\it et al.}}
\def\ibid{{\it ibid}.}
\def\sub#1{_{\lower.25ex\hbox{$\scriptstyle#1$}}}
\def\sul#1{_{\kern-.1em#1}}
\def\sll#1{_{\kern-.2em#1}}
\def\sbl#1{_{\kern-.1em\lower.25ex\hbox{$\scriptstyle#1$}}}
\def\ssb#1{_{\lower.25ex\hbox{$\scriptscriptstyle#1$}}}
\def\sbb#1{_{\lower.4ex\hbox{$\scriptstyle#1$}}}

\def\to{\rightarrow}
\def\mh{\ifmmode m\sbl H \else $m\sbl H$\fi}
\def\mch{\ifmmode m_{H^\pm} \else $m_{H^\pm}$\fi}
\def\mt{\ifmmode m_t\else $m_t$\fi}
\def\mc{\ifmmode m_c\else $m_c$\fi}
\def\mz{\ifmmode M_Z\else $M_Z$\fi}
\def\mw{\ifmmode M_W\else $M_W$\fi}
\def\mws{\ifmmode M_W^2 \else $M_W^2$\fi}
\def\mhs{\ifmmode m_H^2 \else $m_H^2$\fi}
\def\mzs{\ifmmode M_Z^2 \else $M_Z^2$\fi}
\def\mts{\ifmmode m_t^2 \else $m_t^2$\fi}
\def\mcs{\ifmmode m_c^2 \else $m_c^2$\fi}
\def\mchs{\ifmmode m_{H^\pm}^2 \else $m_{H^\pm}^2$\fi}
\def\ztwo{\ifmmode Z_2\else $Z_2$\fi}
\def\zone{\ifmmode Z_1\else $Z_1$\fi}
\def\mtwo{\ifmmode M_2\else $M_2$\fi}
\def\mone{\ifmmode M_1\else $M_1$\fi}
\def\tb{\ifmmode \tan\beta \else $\tan\beta$\fi}
\def\xw{\ifmmode x\sub w\else $x\sub w$\fi}
\def\ch{\ifmmode H^\pm \else $H^\pm$\fi}
\def\lum{\ifmmode {\cal L}\else ${\cal L}$\fi}
\def\inpb{\ifmmode {\rm pb}^{-1}\else ${\rm pb}^{-1}$\fi}
\def\infb{\ifmmode {\rm fb}^{-1}\else ${\rm fb}^{-1}$\fi}
\def\epem{\ifmmode e^+e^-\else $e^+e^-$\fi}
\def\ppb{\ifmmode \bar pp\else $\bar pp$\fi}

\def\bsg{\ifmmode b\rightarrow s\gamma \else $b\rightarrow s\gamma$\fi}

\newskip\zatskip \zatskip=0pt plus0pt minus0pt
\def\matth{\mathsurround=0pt}

\def\atversim#1#2{\lower0.7ex\vbox{\baselineskip\zatskip\lineskip\zatskip
  \lineskiplimit 0pt\ialign{$\matth#1\hfil##\hfil$\crcr#2\crcr\sim\crcr}}}

\renewcommand{\thefootnote}{\fnsymbol{footnote}}

\begin{document} \begin{titlepage}
\setcounter{page}{1}
\thispagestyle{empty}
\rightline{\vbox{\halign{&#\hfil\cr
&SLAC-PUB-6512\cr
&May 1994\cr
&T/E\cr}}}
\vspace{0.8in}
\begin{center}

{\Large\bf
Probing Anomalous Chromomagnetic Top Quark Couplings at the NLC}
\footnote{Work supported by the Department of
Energy, contract DE-AC03-76SF00515.}
\medskip

\normalsize THOMAS. G. RIZZO
\\ \smallskip
{\it {Stanford Linear Accelerator Center\\Stanford University,
Stanford, CA 94309}}\\

\end{center}

\begin{abstract}

The Next Linear Collider(NLC) will provide a excellent tool for probing the
detailed nature of the top quark. By extending the recent analysis of
Dokshitzer, Khoze and Sterling, we perform a preliminary examination of the
influence of an anomalous chromomagnetic moment for the top, $\kappa$,
on the spectrum of gluon radiation associated with $t\bar t$ production. In
particular, we analyze the sensitivity of future data to non-zero values of
$\kappa$ and estimate the limits that can be placed on this parameter at the
NLC with center of mass energies $\sqrt {s}=$ 500 and 1000 GeV.

\end{abstract}

\vskip0.45in
\begin{center}

Submitted to Physical Review {\bf D}.

\end{center}


\renewcommand{\thefootnote}{\arabic{footnote}} \end{titlepage}


Direct searches for the top quark at the Tevatron have led to a lower bound
on the top mass of 131 GeV and, quite recently, to evidence that the top has
been found with a mass near 175 GeV{\cite {cdf,d0}}. If verified, this will be
a remarkable success for the Standard Model(SM) since this value of the mass
lies close to the center of the range predicted by precision electroweak
data{\cite {moriond}}. Due to its large mass, the top itself has been proposed
as a probe for new physics beyond the SM. Detailed analyses of top quark
couplings to gauge bosons through its direct production and subsequent decay
at both hadron{\cite {htop}} and $e^+e^-$ colliders{\cite {etop}} have
been advocated in the literature for this very purpose. The present indirect
constraints from low energy processes still allow for
sizeable deviations from SM predictions{\cite {low}}. In fact, the somewhat
larger than expected cross section for $t\bar t$ production{\cite {qcd}}
obtained by the CDF Collaboration has already prompted several theoretical
analyses{\cite {fermit}} in which new dynamics involving the top quark have
been discussed. Thus it is possible that the top may show us the first glimmer
of new physics beyond the standard model.

The possibility of gluon emission during heavy quark production in $e^+e^-$
collisions has been entertained for quite some time.
In a recent paper, Dokshitzer, Khoze and Sterling(DKS){\cite {dks}} have
considered the spectrum of energetic gluon jets produced in association with
$t\bar t$ at the
Next Linear Collider(NLC). In the present paper we will extend their analysis
and consider the possibility that the top possesses a non-zero anomalous
chromomagnetic dipole moment, $\kappa$, in its coupling to gluons. (As in the
DKS study, we will ignore the effects of top decay in this analysis.) Such a
scenario has recently been shown to lead to significant modifications in
the characteristics of the $t\bar t$ production at the Tevatron
{\cite {david}}. Here in this preliminary analysis we will show that
the gluon energy distribution is a sensitive probe of $\kappa$, as
a non-zero value of this parameter leads to an enhancement in
the number of high energy gluons produced along with $t\bar t$ at the NLC.
Since the emission rate for gluons from
a chromomagnetic dipole term in the Lagrangian scales approximately like
$\kappa\sqrt{s}/m_t$ in the amplitude, large values of $\kappa$ will lead to
a breakdown in perturbation theory. We will also see that for a fixed top quark
mass, increasing the NLC's center of mass energy will provide an additional
lever arm in obtaining sensitivity to non-zero values of $\kappa$.
In the {\it absence} of additional high energy gluon jets in comparison
to ordinary QCD expectations we will demonstrate that reasonable limits
on the value of $\kappa$ are obtainable at the NLC. One might expect that
substantially improved limits from fits to the gluon energy spectrum itself
may be obtainable. We show that vast improvements in the constraints on
$\kappa$ from this approach are unlikely at the $\sqrt {s}=500$ GeV NLC due
to a conspiracy between SM and $\kappa$-dependent contributions to the cross
section. For the $\sqrt {s}=1$ TeV case, however, we find that fits to the
gluon energy distribution yield greatly increased sensitivity to non-zero
$\kappa$.

It is important to note that since the anomalous $t\bar tg$ coupling does
not occur at the primary production vertex at the NLC and appears only in a
higher order process, the limits we expect on the value of
$\kappa$ should be inferior to those obtained in the literature on anomalous
{\it electroweak} top couplings{\cite {etop}}. While this expectation
is realized, the limits we obtain, particularly for the $\sqrt{s}=1$ TeV
$e^+e^-$ collider scenario, are reasonably strong.

It is also important to remember, of course,
that an `anomalous' chromomagnetic moment for the top (or any quark) is
induced at the one-loop level in conventional
QCD and is of order ${\alpha_s\over {\pi}}$. In the context of the present
paper, by `anomalous' we mean a value over and above that given within the SM
context, usually with a magnitude somewhat larger than that induced via
conventional perturbative loop diagrams.

To begin our analysis and in order to set our conventions, we note that the
piece of the Lagrangian which governs the $t\bar t g$ coupling with a non-zero
value of $\kappa$ is given by :
\begin{equation}
{\cal L}=g_s\bar t T_a \left( \gamma_\mu+i{F_2(k^2)\over {2m_t}}\sigma_{\mu\nu}
k^\nu\right)t G_a^\mu \,,
\end{equation}
where $g_s$ and $T_a$ are the usual $SU(3)_c$ coupling and generators, $m_t$
is the top quark mass, $k$ is the out-going gluon momentum, and $F_2$
represents a general, $k^2-$dependent, form factor. (A potential
$F_1(k^2)$-type form-factor has already been set to unity.) For on-shell
gluons, we
define $F_2(k^2=0)=\kappa$ following the standard notation. The rest of the
notation below follows closely that of DKS. Let $p_1$, $p_2$, and $k$ be the
momenta of the $t$, $\bar t$, and $g$ in the final state such that
$q=p_1+p_2+k$ with $q^2=s$. The kinematics of the $e^+e^- \to t\bar t g$
process then imply the usual defining relationships
\begin{eqnarray}
z_i &=& 2q\cdot p_i/s \,, \nonumber \\
z   &=& 2q\cdot k/s  \,, \nonumber \\
2k\cdot p_1/s &=& 1-z_2 \,, \nonumber \\
2k\cdot p_2/s &=& 1-z_1 \,, \nonumber \\
2p_1\cdot p_2/s &=& 1-z-{2m_t^2\over {s}} \,,
\end{eqnarray}
where $z_1+z_2+z=2$ is the statement of energy conservation. Following DKS,
to leading order in $\alpha_s$ we can factorize the weighted, angular
integrated, double
differential cross section for the process of interest into the separate
contributions
due to the vector and axial-vector couplings of the top quark to the s-channel
exchange gauge bosons as
\begin{eqnarray}
{d^2W\over {dz_1dz_2}} &=& F_v{d^2W_v\over {dz_1dz_2}}
+F_a{d^2W_a\over {dz_1dz_2}} \,,
\end{eqnarray}
where $F_{v,a}$ are the `weighting' factors telling us the fraction of events
arising from the vector and axial-vector couplings of the top to $\gamma$
and $Z$. Note that we have scaled our result to the lowest order $t\bar t$
production cross section, \ie, $W=\sigma/\sigma_0$, where
$\sigma_0=\sigma(e^+e^- \to t\bar t)$. (Clearly, we must also have
$F_v+F_a=1$ in the above expression to
conserve probability.) This factorization approach continues to remain
valid even in
the presence of a non-zero $\kappa$. In order to proceed, only the quantities
${d^2W_{v,a}\over {dz_1dz_2}}$ and $F_{v,a}$ need to be computed. Since
$F_{v,a}$ are insensitive to the existence of the anomalous chromomagnetic
moment of the top quark, they are given solely by the kinematics and the
electroweak couplings of the top. These factors can be read off directly from
the general cross section expression for the production of heavy fermion pairs
in $e^+e^-$ collisions, \eg, in{\cite {pr}}:
\begin{eqnarray}
F_v &=& {{1\over {2}}\beta(3-\beta^2)A_v\over {\beta^3 A_a+
{1\over {2}}\beta(3-\beta^2)A_v}} \,, \nonumber \\
F_a &=& {\beta^3 A_a\over {\beta^3 A_a+{1\over {2}}\beta(3-\beta^2)A_v}} \,,
\end{eqnarray}
and $\beta=\sqrt{(1-4m_t^2/s)}$, with $s$ being the square
of the center of mass energy. $A_{v,a}$ are directly determined by the vector
and axial vector couplings of the electron and top quark as well as the gauge
boson propagator function:
\begin{eqnarray}
A_v &=& \sum_{ij} ~(v_iv_j+a_ia_j)_e(v_iv_j)_t P_{ij} \,, \nonumber \\
A_a &=& \sum_{ij} ~(v_iv_j+a_ia_j)_e(a_ia_j)_t P_{ij} \,, \nonumber \\
P_{ij} &=& s^2 {[(s-M_i^2)(s-M_j^2)+(\Gamma M)_i(\Gamma M)_j]
\over {[(s-M_i^2)^2+(\Gamma M)_i^2][(s-M_j^2)^2+(\Gamma M)_j^2]}} \,.
\end{eqnarray}
The sum in the expression above is over the s-channel $\gamma(i,j=1)$ and
$Z(i,j=2)$ gauge boson exchanges, including finite width effects, and for
numerical purposes in our analysis the couplings are normalized
to the running electromagnetic charge. In these numerical calculations, we
will use the values of the various parameters as given
in Ref.{\cite {moriond}}.
We assume that the vector and axial vector couplings of the top quark are
given by their conventional SM values with no alterations being present in the
$t\bar t  \gamma,Z$ vertices.

Given all of the above, the evaluation of the square of the matrix element is
quite straightforward. Defining the overall normalization coefficients
\begin{eqnarray}
N_v &=& {2\alpha_s\over {3\pi}}(2m_t^2s^2x_1^2x_2^2)^{-1}[{1\over{2}}\beta
(3-\beta^2)]^{-1} \,, \nonumber \\
N_a &=& {2\alpha_s\over {3\pi}}(2m_t^2s^2x_1^2x_2^2)^{-1}[\beta^3]^{-1} \,,
\end{eqnarray}
where $x_i=1-z_i$, we obtain the following expressions for the above
distributions:
\begin{eqnarray}
{d^2W_v\over {dz_1dz_2}} &=& N_v\left[ -8m_t^6(x_1+x_2)^2-4sm_t^4
[x_1^2(1+2x_2)+x_2^2(1+2x_1)]+2s^2m_t^2x_1x_2 \right. \nonumber \\
 & & \left. [(1-x_1)^2+(1-x_2)^2+\kappa(x_1-x_2)^2]+\kappa^2s^3x_1^2x_2^2
(1-x_1-x_2)\right] \,, \nonumber \\
{d^2W_a\over {dz_1dz_2}} &=& N_a\left[ 16m_t^6(x_1+x_2)^2+2sm_t^4
[(\kappa^2+2\kappa+2)x_1x_2(x_1+x_2)^2+8x_1x_2 \right.  \nonumber \\
 & & (x_1+x_2)-2(x_1^2+x_2^2+6x_1x_2)]+2m_t^2s^2x_1x_2[(1-x_1)^2+(1-x_2)^2+
 \nonumber \\
 & & \left. \kappa(x_1^2+x_2^2-4)+\kappa^2x_1x_2(x_1+x_2-3)]+
\kappa^2s^3x_1^2x_2^2(1-x_1)(1-x_2)\right] \,,
\end{eqnarray}
which, when combined with the other results above, gives the
complete $t\bar t g$ double
differential cross section, normalized to that for $t\bar t$ production,
including the contributions from finite $\kappa$. For $\kappa=0$, we
reproduce the standard results in the literature{\cite {dks}}. These
expressions have no co-linear singularities, due to the finite top mass, but
are still, overall, infrared singular in the limit of zero gluon momenta.
It is already
clear from Eqs. 6 and 7 that as $s/m_t^2$ gets large the last terms in the
expressions for ${d^2W_{v,a}\over {dz_1dz_2}}$ will become dominant over the
conventional QCD result. Note that these terms are infrared finite due to the
fact that the chromomagnetic coupling in the Lagrangian is proportional to the
gluon momenta. Semiquantitative bounds on $\kappa$ follow immediately upon
integrating these two terms over phase space thus obtaining the ratio of the
$t\bar tg$ to $t\bar t$ cross sections in the large $s/m_t^2$ limit, \ie,
\begin{eqnarray}
{\sigma_{t\bar tg}\over {\sigma_{t\bar t}}} &\simeq& {\alpha_s\over {\pi}}
{\kappa^2s\over {18m_t^2}} {v^2+1.25a^2\over {v^2+a^2}}\,,
\end{eqnarray}
which leads for $m_t=175$ GeV and $\sqrt {s}=500$ GeV ( assuming that the
large $s/m_t^2$ limit is crudely valid in this case) to
$\simeq 5.0(\kappa/3)^2$
as the coefficient of ${\alpha_s\over {\pi}}$. Correcting for a color
factor and a missing ${1\over {\pi}}$, this limiting behavior is in
agreement with that
obtained by Grifols and Mendez{\cite {gm}} for an anomalous magnetic moment
of the $\tau$ in the decay $Z \to \tau\bar \tau \gamma$. This crude
estimate tells us
that for perturbation theory to make sense for $t\bar t g$ production at the
NLC, the value of $\kappa$ must approximately satisfy $|\kappa| \leq 3$.

There are several ways to see the effects of a non-zero $\kappa$ on the gluon
jet energy spectrum. For the moment, let us follow DKS and calculate the
average value of the scaled gluon energy, \ie, $z_{ave}$,
(where $z=2E_{glu}/\sqrt {s}$) as a function of $\beta^2$ for pure vector or
pure axial vector couplings; these are shown in Figs. 1a and 1b (assuming
$\alpha_s=0.1$ for purposes of direct comparison with DKS). Note that this
quantity, $z_{ave}$, is infrared safe for all values of $\beta$.
For $\kappa=0$, the results of DKS are immediately recovered but for non-zero
$\kappa$, significant deviations are observed which grow quite large with
increasing $\beta^2$. It is, of course, just in this phase space regime where
we expect the largest deviations from the conventional QCD results
since large $\beta^2$ implies large $s/m_t^2$. Of course, even in standard
QCD, the region close to $\beta=1$ becomes non-perturbative (unless further
cuts are applied) since it
corresponds to the location in phase space where multiple soft co-linear gluon
emission can occur. However, long before this non-perturbative range is
reached we see that for non-zero values of
$\kappa$ there is an upward shift in the average value of $z$, for both
vector and axial vector couplings. This increased divergence in the
average value of
$z$ for large $\beta^2$ is only symptomatic of a more widespread phenomena,
\ie, the the presence of a non-zero $\kappa$ not only increases the rate for
the $t\bar t g$ final state but also hardens the gluon jet energy
distribution. We further note that the region of small $\beta^2$ is also
nonperturbative as this corresponds to the threshold regime where
resummation techniques need to be applied. The possible
signatures for a non-zero $\kappa$ from detailed threshold region studies is
beyond the scope of the present paper.

Perhaps a better probe of non-zero $\kappa$ is the gluon jet energy spectrum
itself. This we show in Fig. 2a assuming $m_t=175$ GeV and an NLC center of
mass energy of $\sqrt {s}=500$ GeV. From the kinematics, the maximum gluon
energy for these input parameters is $z_{max}=1-4m_t^2/s=0.51$ or 127.5 GeV.
Note that in the Figure we have again assumed $\alpha_s=0.1$. To get the
`best' result to compare with experiment (ignoring higher order corrections)
we need to obtain the correct value of $\alpha_s$ at this energy and simply
rescale all of the curves in Fig. 2a by an overall factor. One possible
approach, which we will follow below, is to use the
method of Brodsky, LePage, and Mackenzie(BLM){\cite {blm}} in order to set
the scale, using the value of $\alpha_s(M_z)${\cite {moriond}} as input, and
to make use of the three-loop renormalization group equations. Without doing a
complete calculation, it is possible to estimate the correct BLM scale($Q^*$)
at which to evaluate $\alpha_s$ by simple phase space considerations
{\cite {stan}}. One finds $Q^*\simeq 0.435(\sqrt {s}-2m_t)/3$, not too
different than the $Q^*$ for $Z$ decay studies, so that we arrive at
$\alpha_s(Q^*)=0.121$ as our estimate (using the value of $\alpha_s$ at the
$Z$ scale extracted from $R_h${\cite {moriond}} as input). This implies
that all the curves in
Fig. 2 should be scaled upward by about $21\%$. For other preferred values of
$\alpha_s$, an appropriate rescaling should be performed.

Of course, the overall scale is not
the interesting part of these figures as far as $\kappa$ sensitivity is
concerned since $\kappa$ non-zero results in a distortion in the gluon
spectrum, especially at larger $z$ values. Generally one sees that the effect
of $\kappa$ is to flatten the spectrum so that there is an excess of gluon
jets with high energies. For $\kappa=-1$, however, we see a sharper fall off
in the spectrum than in standard QCD; this is a result of a destructive
interference between the ordinary and $\kappa$-dependent amplitudes which
takes place for the specific values of $m_t$ and $\sqrt {s}$ we have used as
input. This implies that for a range of negative $\kappa$, the SM result and
the one where $\kappa$ is non-zero will be very difficult to distinguish.
Since the deviation due to $\kappa$ is at larger $z$ values and the
spectrum diverges as $z \to 0$, we can apply a cut on the minimum gluon energy,
$z_{cut}$, which we use to define our event sample and integrate the
spectrum for
gluon energies above that value. This results in Fig. 2b, which should also
be scaled upwards by $21\%$ if the BLM approach is used. As expected, the
curves for non-zero $\kappa$ are generally higher than the standard
QCD result. Taking the BLM value for $\alpha_s$ and a value of $z_{cut}=0.2$,
\ie, only events with gluon jets having energies in excess of 50 GeV, we show
in Fig. 2c the $\kappa$ dependence of the resulting integrated cross section.
Assuming an integrated luminosity of $30 fb^{-1}$, this corresponds to a
sample of $375(I/0.02)$ $t\bar t g$ events before further cuts are applied
or 457 events in the standard QCD case with $\kappa=0$. Of
course to identify top pair production, we will demand at least one high-$p_t$
lepton in the event($B=0.44$) and, perhaps, an additional $b$-tag(with an
assumed efficiency of $\epsilon=0.8$) to remove backgrounds.

If the SM result is realized, we can use the estimates of the $t\bar tg$
event rate above to place bounds on the value of $\kappa$. (We might expect
better limits
would most likely be obtainable by a direct measurement of the gluon energy
spectrum instead of a simple rate estimate; we will return to this possibility
below.) Allowing for a $2\%$ error on the determination of
$\alpha_s$ in the NLC era, a $5\%$ systematic error from higher-order QCD
uncertainties, and the rates above to calculate statistical errors, the value
of $I$ would be determined to be $2.44\pm 0.23\%$, which at $95\%$ CL would
restrict $\kappa$ to lie in the range $-2.1 \leq \kappa \leq 0.6$. Varying
the cuts leads to numerically similar results there being a relative trade
off between increased(decreased) statistics and decreased(increased)
sensitivity.  The reason for the poor limit is clear; as we saw above a range
of negative $\kappa$ exists for which the resulting energy distribution is
almost identical to the SM one. This is a result of a conspiracy between the
various contributions to the differential cross section and the particular
values of $m_t$ and $\sqrt {s}$ we are examining.

This situation is not much alleviated by an actual fit
to the gluon energy distribution itself. For $z \geq 0.15$, we generate
`data' assuming $\kappa=0$ by Monte Carlo taking a bin size of
$\Delta z=0.05$.  The phase space region of interest is then covered by a
total of seven $z$ bins, the last one covering the range
$0.45 \leq z \leq \beta^2=0.51$. We then fit the $\kappa$-dependent gluon
spectrum to the data and perform a $\chi^2$ analysis. The resulting $\chi^2$
is a quartic function of $\kappa$ and is shown explicitly in Fig. 2d where one
sees that two essentially degenerate minima exists. This is due to the
conspiracy discussed above and, as a result, only rather poor bounds on
$\kappa$ are obtainable. Explicitly, from this procedure we obtain the
allowed range $-1.98 \leq \kappa \leq 0.44$
at $95\%$ CL, which is only a slight improvement in the limit obtained above
from simple counting. Similar limits are obtained if different bin sizes are
chosen for the Monte Carlo study but we have not tried to optimize this choice
in our analysis. In Fig. 2e, we compare the data generated by Monte Carlo with
the energy spectra predicted for the $\kappa$ values corresponding to the two
approximately degenerate $\chi^2$ minima, \ie, $\kappa=-1.69$ and $0.12$. As
can be seen the fit is quite good in both cases.

What happens when we go to larger values of $\sqrt {s}$ where significantly
greater sensitivities to $\kappa$ are expected? In Figs. 3a-d, we examine the
case of $t\bar t g$ production with non-zero $\kappa$ for an NLC with
$\sqrt {s}=1$ TeV.  (In these figures, the BLM value of $\alpha_s=0.100$ has
been assumed following the above procedure so that no overall rescaling is
necessary in this case.) Fig. 3a clearly shows that at large values of $z$,
the gluon jet energy distribution is  even more enhanced for fixed values of
$\kappa$ than for the $\sqrt {s}=500$ GeV case; this is exactly what we should
have expected.  Note the approximate symmetry
of the curves under the interchange $\kappa \to -\kappa$; this occurs
naturally in the large $s/m_t^2$ limit as seen above. Applying the $z_{cut}$
approach as before yields Fig. 3b where the $\kappa \to -\kappa$ symmetry is
even more obvious. The value of $I$ is so large in the $|\kappa|=3$ case for
values of $z_{cut}<0.4$ that we should most likely not trust our lowest order
perturbative result for this range of parameters. Taking $z_{cut}=0.4$, which
corresponds to a gluon jet of energy 200 GeV, we
show in Fig. 3c the explicit $\kappa$ dependence of our cross section ratio.
If we assume an integrated luminosity of $200 fb^{-1}$ and make the
same assumptions as in the
500 GeV case, the realization of the SM result can again be used to place
significant constraints on $\kappa$. In this case we would obtain
$I=5.24\pm0.36 \%$, which at $95\%$ CL would restrict $\kappa$ to lie in
the range $-1.0 \leq \kappa \leq 0.25$, which is about a factor of two better
than that obtained for the 500 GeV NLC. We see again that because small,
negative values of $\kappa$ and the SM case lead to very similar gluon energy
distributions, we do not obtain a very narrow allowed range for $\kappa$.
However, if we try the Monte Carlo approach as we did above and
fit the entire spectrum for $z \geq 0.4$ in nine $z$ bins (the last bin
covering the range $0.8 \leq z \leq \beta^2=0.8775$) we obtain the $\chi^2$
plot shown in Fig. 3d. Unlike the $\sqrt {s}=500$ GeV case, the second local
$\chi^2$ minima is no longer degenerate so that we can now obtain a
substantially improved bound on $\kappa$: $-0.12 \leq \kappa \leq 0.21$.
Again, we have not made any attempt to optimize the bin size in this Monte
Carlo study. In Fig. 3e, we compare the Monte Carlo generated data with the
predicted gluon spectrum for the choice of $\kappa=0.06$, corresponding to
the $\chi^2$ minimum. As can be seen, the fit is quite good.

In this paper we have analyzed the influence of an anomalous chromomagnetic
dipole moment for the top quark, $\kappa$, on the cross sections
and associated gluon jet energy distributions for $t\bar tg$ events produced
at both  500 and 1000 GeV $e^+e^-$ linear colliders assuming a top quark
mass of 175 GeV. Making a cut on the gluon jet energy of 50(200) GeV and
demanding at least one b-tag as well as one high-$p_t$ lepton tag as a top
signal, an integrated luminosity of $30(200) fb^{-1}$ leads to a bound on
$\kappa$ of $-2.1 \leq \kappa \leq 0.6(-1.0 \leq \kappa \leq 0.25)$ at the
NLC with $\sqrt{s}=500(1000)$ GeV, assuming no excess $t\bar tg$ events
are observed. One might have expected that a complete fit of the
gluon energy spectrum to the $\kappa$-dependent distribution would generally
lead to
substantial improvements in these limits. We found, however, that this was
not the case at a $\sqrt {s}=500$ GeV NLC due to a conspiracy between the SM
and $\kappa$-dependent terms in the cross section and that only slight
improvements were obtainable: $-1.98 \leq \kappa \leq 0.44$. For the case of
a $\sqrt{s}=1$ TeV machine, this conspiracy no longer took place and the power
of fitting to the gluon energy spectrum was realized yielding a vastly
improved bound of $-0.12 \leq \kappa \leq 0.21$.

The results of this analysis are only preliminary.
In a more complete Monte Carlo study the effects of the top decay,
possible gluon emission from the final state bottom quarks, and detector
resolution and efficiencies need to be included.
However, by demanding a very high energy additional gluon jet
with the rest of the event reconstructing to $t\bar t$, the results of such
an analysis should closely mimic those we have obtained above. We have seen
that it is quite likely that the NLC will be able to place reasonably strong
constraints on the existence of a top quark anomalous chromomagnetic moment.
Of course, the more exciting possibility of
observing  a hardening of the gluon jet energy spectrum would be a spectacular
signature for an anomalous magnetic moment for the top or other  new physics
beyond the standard model.

\vskip.25in
\centerline{ACKNOWLEDGEMENTS}

The author like to thanks A. Kagan, D. Atwood, J.L. Hewett, L. Dixon,
T. Barklow, P. Burrows and S. Brodsky for discussions related to this work.
He would also like to thank the members of the Argonne National Laboratory
High Energy Theory Group for use of their computing facilities.

\newpage

%
\def\MPL #1 #2 #3 {Mod.~Phys.~Lett.~{\bf#1},\ #2 (#3)}
\def\NPB #1 #2 #3 {Nucl.~Phys.~{\bf#1},\ #2 (#3)}
\def\PLB #1 #2 #3 {Phys.~Lett.~{\bf#1},\ #2 (#3)}
\def\PR #1 #2 #3 {Phys.~Rep.~{\bf#1},\ #2 (#3)}
\def\PRD #1 #2 #3 {Phys.~Rev.~{\bf#1},\ #2 (#3)}
\def\PRL #1 #2 #3 {Phys.~Rev.~Lett.~{\bf#1},\ #2 (#3)}
\def\RMP #1 #2 #3 {Rev.~Mod.~Phys.~{\bf#1},\ #2 (#3)}
\def\ZP #1 #2 #3 {Z.~Phys.~{\bf#1},\ #2 (#3)}
\def\IJMP #1 #2 #3 {Int.~J.~Mod.~Phys.~{\bf#1},\ #2 (#3)}

\newpage

%
{\bf Figure Captions}
\begin{itemize}

\item[Figure 1.]{Average value of the scaled gluon jet energy, $z$, as a
function of $\beta^2$ for both purely vector(a) or axial vector(b) heavy quark
couplings. The upper(lower) dotted, dashed, and dot-dashed curves correspond
to $\kappa$ values of 3(-3), 2(-2), and 1(-1) respectively while the solid
curve is conventional QCD with $\kappa=0$. $\alpha_s=0.10$ has been assumed.}
\item[Figure 2.]{(a)Gluon jet energy spectrum assuming $\alpha_s=0.10$ for
$m_t=175$ GeV at a $\sqrt {s}=500$ GeV NLC. (b)Integrated gluon energy
spectrum for the same input parameters as in (a) as a function of the minimum
gluon energy, $z_{cut}$. In both (a) and (b), the labelling of the various
curves is as in Fig.~1. (c) Same as (b) but as a function of $\kappa$ assuming
$z_{cut}=0.2$ and $\alpha_s=0.121$ as suggested by the BLM approach. (d)
$\chi^2$ fit to the gluon energy spectrum for the value of $\kappa$ as
described in the text. $\alpha_s=0.121$ is again assumed. (e) Best fit gluon
$\kappa$-dependent spectra through the points generated by Monte Carlo. The
dashed(dash-dotted) curve corresponds to to $\kappa=-1.69(0.12)$.}
\item[Figure 3.]{Same as Fig.~2, but for an NLC with $\sqrt {s}=1$ TeV. In (c)
the value of $z_{cut}=0.4$ is assumed along with $\alpha_s=0.100$. (e) Best
fit gluon spectrum through the points generated by the Monte Carlo analysis
corresponding to $\kappa=0.06$.}
\end{itemize}


\begin{thebibliography}{99}
\bibitem{cdf}
F.\ Abe \etal, CDF Collaboration, Fermilab report Fermilab-PUB-94/097-E, 1994.
\bibitem{d0}
S.\ Abachi \etal, D0 Collaboration, \PRL 72 2138 1994 .
\bibitem{moriond}
See talks given by P.\ Siegrist, D.\ Bardin, P.\ Clarke and B.\ Pietrzyk
at the {\it XXIXth Recontres de Moriond, Electroweak Interactions and Unified
Theories}, Meribel, France, March 1994.
\bibitem{htop}
G.\ Kane, G.A.\ Ladinsky and C.P.\ Yuan, \PRD D45 124 1992 ;
C.P.\ Yuan, \PRD D45 782 1992 ;
D.\ Atwood, A.\ Aeppli and A.\ Soni, \PRL 69 2754 1992 ;
D.\ Atwood, A.\ Kagan and T.G.\ Rizzo, SLAC report (in preparation).
\bibitem{etop}
M.\ Peskin, talk presented at the {\it Second International Workshop on Physics
and Experiments at Linear $e^+e^-$ Collider},  Waikoloa, HI, April 1993;
M.\ Peskin and C.R.\ Schmidt, talk presented at the {\it First Workshop on
Linear Colliders}, Saariselk\" a, Finland, September 1991; P.\ Zerwas, \ibid;
W.\ Bernreuther \etal, in {\it Proceedings of
the Workshop on \epem\ Collisions at 500 GeV, The Physics Potential},
(DESY, Hamburg) ed.\ by P.\ Igo-Kemenes and J.H.\ K\" uhn, 1992; A.\ Djouadi,
ENSLAPP-A-365-92 (1992);
M.\ Frigeni and R.\ Rattazzi, \PLB B269 412 1991 ;
R.D.\ Peccei, S.\ Persis and X.\ Zhang, \NPB B349 305 1991 ;
D.O.\ Carlson, E.\ Malkawi and C.-P.\ Yuan, Michigan State report
MSUHEP-94/05, 1994.
\bibitem{low}
Limits on $\kappa$ from the decay $b \to s \gamma$ were considered by
J.L.\ Hewett and T.G.\ Rizzo, \PRD D49  319 1994 .
\bibitem{qcd}
For the most recent estimate of the top production cross section at
next-to-leading order including gluon resummation in QCD, see E.\ Laenen,
J.\ Smith, and W.L.\ van Neerven, \PLB B321 254 1994 .
\bibitem{fermit}
See, for example, C.\ Hill and S.\ Parke, Fermilab report
Fermilab-PUB-93/397-T, 1993; E.\ Eichten and K.\ Lane, Fermilab report
Fermilab-PUB-94/007-T, 1994; see also, V.\ Barger and R.J.N.\ Phillips,
University of Wisconsin report MAD/PH/830, 1994.
\bibitem{dks}
Yu.L.\ Dokshitzer, V.A.\ Khoze and W.J.\ Sterling, University of Durham
report DTP/94/14, 1994. The production of $q\bar q g$ final states for
heavy quarks in both $Z$ decay and $e^+e^-$ annihilation were first
considered long ago; see, \eg, B.L.\ Ioffe, \PLB B78 277 1978 ;
G.\ Kramer, G.\ Schierholz and J.\ Willrodt, \ZP C4 149 1980 ;
E.\ Learman and P.M.\ Zerwas, \PLB B89 225 1980 ;
G.\ Grunberg, Y.J.\ Ng and S.-H.H.\ Tye, \PRD D21 62 1980 ;
T.R.\ Taylor, \ZP C2 313 1979 ; H.P.\ Nilles, \PRL 45 319 1980 ;
T.G.\ Rizzo, \PRD D22 2213 1980 .
\bibitem{david}
See D.\ Atwood, A.\ Kagan, and T.G.\ Rizzo in Ref.{\cite {htop}}.
\bibitem{pr}
See, for example, Eq. 3.101 in J.L.\ Hewett and T.G.\ Rizzo, \PR 185 193 1989 .
\bibitem{gm}
A.\ Grifols and A.\ Mendez, \PLB B255 611 1991 .
\bibitem{blm}
S.\ Brodsky, P.\ LePage and P.B.\ Mackenzie, \PRD D28 228 1983 ;
S.\ Brodsky and H.J.\ Lu, SLAC report SLAC-PUB-6481, 1994.
\bibitem{stan}
The author would like to thank Stan Brodsky for this helpful suggestion.


\end{thebibliography}
\end{document}